\documentclass[
  journal=pasa,
  manuscript=research-article, 
  year=2024
]{cup-journal}

\usepackage{ulem}
\usepackage{orcidlink}
\usepackage{bm}
\usepackage[utf8]{inputenc}
\usepackage{amsmath}
\usepackage[nopatch]{microtype}
\usepackage{booktabs}
\usepackage{hyperref} 
\hypersetup{colorlinks,citecolor=blue,linkcolor=blue,urlcolor=blue}

\usepackage{caption} 
\usepackage{amsfonts}
\usepackage{amssymb} 
\usepackage[pagewise]{lineno}

\def\v2{V_2\left(\mathbf{U},\Delta \nu = 0\right)}

\def\V{\mathcal{V}}

\newcommand{\HI}{{\rm H\hspace{0.5mm}}{\scriptsize {\rm I}}} 

\newcommand{\dnu}{\Delta\nu}

\newcommand{\vtht}{\bm{\theta}}

\newcommand{\eg}{{\it e.g.}\,}

\title[Angular power spectrum of the DGSE]{A measurement of Galactic synchrotron emission using MWA drift scan observations}

\author{Suman Chatterjee\,\orcidlink{0000-0001-8852-5888}}
\affiliation{Department of Physics and Astronomy, University of the Western Cape,
7535 Bellville, Cape Town, South Africa}

\author{Shouvik Sarkar\,\orcidlink{0009-0003-3096-7028}}
\affiliation{Centre for Strings, Gravitation and Cosmology, Department of Physics, Indian Institute of Technology Madras, Chennai 600036, India}

\author{Samir Choudhuri}
\affiliation{Centre for Strings, Gravitation and Cosmology, Department of Physics, Indian Institute of Technology Madras, Chennai 600036, India}
\email[Samir Choudhuri]{samir@iitm.ac.in}

\author{Khandakar Md Asif Elahi\,\orcidlink{0000-0003-1206-8689}}
\affiliation{Centre for Strings, Gravitation and Cosmology, Department of Physics, Indian Institute of Technology Madras, Chennai 600036, India}
\alsoaffiliation{Department of Physics, Indian Institute of Technology Kharagpur, Kharagpur - 721 302, India.}

\author{Somnath Bharadwaj\,\orcidlink{0000-0002-2350-3669}}
\affiliation{Department of Physics, Indian Institute of Technology Kharagpur, Kharagpur - 721 302, India.}

\author{Shiv Sethi}
\affiliation{Raman Research Institute, C. V. Raman Avenue, Sadashivanagar, Bengaluru 560080, India.}

\author{Akash Kumar Patwa\,\orcidlink{0000-0002-6216-2430}}
\affiliation{Raman Research Institute, C. V. Raman Avenue, Sadashivanagar, Bengaluru 560080, India.}

\keywords{large-scale structure of universe--first stars--cosmology:reionization--diffuse radiation, methods: statistical, technique--interferometric.}

\begin{document}

\begin{abstract}

Studying the diffuse Galactic synchrotron emission (hereafter, DGSE) at arc-minute angular scale is important to remove the foregrounds for the cosmological 21-cm observations. Statistical measurements of the large-scale DGSE can also be used to constrain the magnetic field and the cosmic ray electron density of our Galaxy's interstellar medium (ISM). Here, we have used the Murchison Widefield Array (MWA) drift scan observations at $154.2 \, {\rm MHz}$ to measure the angular power spectrum $({\cal C}_{\ell})$ of the DGSE of a region of the sky from right ascension (RA) $349^{\circ}$ to $70.3^{\circ}$ at the fixed declination $-26.7^{\circ}$. In this RA range, we have chosen 24 pointing centers (PCs), for which we have removed all the bright point sources above $\sim430 \, {\rm mJy}\,(3\sigma)$, and applied the Tapered Gridded Estimator (TGE) on residual data to estimate the ${\cal C}_{\ell}$. We use the angular multipole range $65 \le \ell \le 650$ to fit the data with a model,  ${\cal C}^M_{\ell}=A\times \left(\frac{1000}{\ell}\right)^{\beta}+C$, where we interpret the model as the combination of a power law $(\propto \ell^{-\beta})$ nature of the DGSE and a constant part due to the Poisson fluctuations of the residual point sources. We are able to fit the model ${\cal C}^M_{\ell}$ for six PCs centered at $\alpha=352.5^{\circ}, 353^{\circ}, 357^{\circ}, 4.5^{\circ}, 4^{\circ}$ and $1^{\circ}$. We run the Markov Chain Monte Carlo (MCMC) ensemble sampler to get the best-fit values of the parameters $A, \beta$ and $C$ for these PCs. We see that the values of $A$ vary in the range $155$ to $400$ mK$^{2}$, whereas the $\beta$ varies in the range $0.9$ to $1.7$. We find that the value of $\beta$ is consistent at $2-\sigma$ level with the earlier measurement of the DGSE at similar frequency and angular scales.   
\end{abstract}

\section{Introduction}
\label{sec:intro}
The epoch of reionisation (EoR), when the neutral hydrogen (\HI) in the inter-galactic medium (IGM) was nearly completely ionised by the first luminous sources, is one of the least known epochs in cosmology.  Direct observations of the EoR using the redshifted \HI{} 21-cm line hold the potential to reveal a substantial volume of astrophysical and cosmological information. Several current and future radio interferometers  aim to measure the power spectrum of  intensity fluctuations of the EoR 21-cm signal, namely  the   Giant Metrewave Radio Telescope (GMRT; \citealt{Swarup1991, Gupta2017}), the Murchison Widefield Array (MWA; 
\citealt{Tingay2013}), the LOw Frequency ARray (LOFAR; \citealt{vanHarlem2013}), the Hydrogen Epoch of Reionization Array (HERA; \citealt{Deboer2017}), and the upcoming SKA-low \citep{Mellema2013, Koopmans2015}.

Measuring the EoR 21-cm signal is particularly challenging due to the presence of strong foregrounds, which are 4-5 orders of magnitude brighter than the expected 21-cm signal \citep{Ali2008, Bernardi2009, Ghosh2012, Paciga2013, Patil2017}. Extra-galactic point sources are the most dominant foreground component at small angular scales,  whereas the DGSE dominates at large angular scales ($ > 10~ {\rm arcmin}$). To mitigate galactic and extra-galactic foreground contamination, interferometric experiments use either foreground avoidance or foreground removal techniques. The foreground avoidance technique relies upon the fact that the foregrounds are intrinsically smooth in frequency and expected to remain restricted in the `foreground wedge' \citep{Datta2010, Thyagarajan2013}. Whereas removal of the foregrounds involves modeling contributions from each component and subtracting them from the observed data. Possibly the most optimal way to extract the EoR 21-cm signal is to use foreground removal in conjunction with the avoidance \citep{Barry2019, Trott2020}. Essentially, these subtraction techniques rely upon modeling the bright compact sources using longer baselines. The EoR 21-cm signal is pronounced in short baselines that remain dominated by the DGSE. Also, incomplete sky models used in the calibration can also lead to the suppression of DGSE in-turn suppressing the EoR 21-cm signal \citep{Byrne2019}. These make it crucial to measure and model the DGSE particularly at small baselines (large angular scales). Further exclusion of these small baselines in the calibration steps can possibly lead to the problem of `excess variance' that is seen in the 21-cm power spectrum estimates \citep{Barry2016}. DGSE models can be used for better calibration thus mitigating these issues. The foreground removal at large scales employs techniques such as Fast Independent Component Analysis (FastICA; \citealt{Maino2002}), Generalized Morphological Component Analysis  (GMCA; \citealt{Bobin2007}), Smooth Component Filtering (SCF; \citealt{Elahi2025}), Gaussian Process Regression (GPR; \citealt{Mertens2020, Elahi2023b}); primarily relies upon the fact that the DGSE is smooth in frequency. An accurate measurement of angular power spectrum $(C_{\ell})$ at different frequencies can quantify the degree of smoothness. Furthermore, current analysis of LOFAR-EoR observations suggests the idea of differentiating the contaminate subtraction process over different distinct spatial scales \citep{Hothi2021}. A measurement of DGSE amplitude at different scales can provide the expected level of contamination to accurate subtraction of foregrounds with relatively lower signal loss. Several 21-cm experiments such as MWA \citep{Byrne2022},OVRO-LWA \citep{Eastwood2018}, and the LWA New Mexico station \citep{Dowell2017} are already being used to develop diffuse sky maps to facilitate the precise calibration of 21-cm experiments.

It is, therefore, of considerable interest to measure and quantify the statistical properties of the DGSE, which is an important foreground while measuring the EoR 21-cm power spectrum. The study of the DGSE is also important in its own right. The Galactic synchrotron radiation is mainly emitted by the relativistic electrons rotating in the magnetic fields. The observed fluctuations of the DGSE at different scales will depend on the fluctuation of both density and magnetic field strength. Also, the magnetohydrodynamic turbulence in the interstellar medium plays a significant role in the observed structures of synchrotron emission. Thus, the $C_{\ell}$ of the DGSE can probe statistics of the density and magnetic field fluctuation as well as about the nature of the turbulence in the plasma \citep{Cho2010,Lazarian2012,Iacobelli2013b}. The largest linear scale of turbulent component of the galactic magnetic field $L_{\rm out }$ that quantifies the scale of energy injection can be used to investigate the interplay between the magnetic field with the turbulence in the ISM. A measurement of the DGSE angular power spectrum can be used to constrain the outer scales  of the turbulence $L_{\rm out }$ and in-turn the relative strength of the magnetic field \citep{Iacobelli2013b}. Also, the observed DGSE can be used to differentiate the contribution in the diffuse emission from the thermal and non-thermal components.

There are several observations spanning a wide range of frequencies which characterise the Galactic synchrotron emission at different angular scales. \citet{Haslam1981} have first measured the brightness temperature of the Galactic synchrotron radiation at 408 MHz radio frequency. Later, \cite{Remazeilles2015} reprocessed the raw data to produce an improved 408-MHz all-sky map. \citet{Reich1982,Reich1986} generated the synchrotron map for the northern sky but at a relatively higher frequency (1.4 GHz). \citet{Reich2001} repeated the analysis for the southern sky using a 30 m radio telescope at Villa Elisa, Argentina. The all-sky spectral index of the synchrotron emission can be measured using these observations at different frequencies \citep{Reich1988, Guzman2011}. \citet{deOliveira-Costa2008} have produced a Global Sky Model (hereafter, GSM) for the synchrotron map using 11 most accurate data in the frequency range 10 MHz to 94 GHz. \citet{Zheng2017} have improved the GSM map by using 29 sky maps from 10 MHz to 5 THz. Upcoming 310 MHz observation using the Green Bank telescope (GBT) along with custom instrumentation expected to produce absolute DGSE map calibrated zero level \citep{Singal2023}.

It is useful to consider the $C_{\ell}$ to quantify 
the two-point statistics of the DGSE.   Several authors have used the all-sky maps to estimate $C_{\ell}$ of the DGSE. At $2.4$ GHz,  \citet{Giardino2001} have found  that   $C_{\ell}$  follows a power-law $C_{\ell} \propto  \ell^{-\beta}$ with $\beta = 2.43 \pm 0.01$ in the $\ell$ range  $2 - 100$, when estimated across the entire sky. They also found that the slope appears to steepen $(\beta=2.92 \pm 0.07)$ at higher Galactic latitudes. \citet{Bennett2003a} have used the {\it Wilkinson Microwave Anisotropy Probe} (WMAP) data and found that $\beta=2$ for the $\ell$ range  $2 -100$. \citet{LaPorta2008} have used data from two different frequencies, 408 MHz and 1.4 GHz, to estimate the $C_{\ell}$ separately at different parts of the sky, for which the values of $\beta$ are found to be in the range $2.6$ to $3.0$ for $\ell<300$ (angular scale greater than 1 deg). However, the angular ranges and the frequencies in most of these observations are larger than those corresponding to most  EoR 21-cm  observations. 

Directly addressing radio-interferometric observations at the angular scales and frequencies relevant for EoR 21-cm  observations, \citet{Bernardi2009} have first analysed Westerbork Synthesis Radio Telescope (WSRT) data to measure  $C_{\ell}$ of the DGSE at 150 MHz for a particular pointing direction.   They found that the power law behaviour $C_{\ell} \propto  \ell^{-\beta}$, and obtained  $\beta=2.2$  at  the angular multipoles $\ell<900$.  \citet{Ghosh2012} have analysed a single pointing of the GMRT 150 MHz observations and found the value $\beta=2.34$ for the $\ell$ range $253 \le \ell \le 800$. \citet{Iacobelli2013a} showed that the $C_{\ell}$ follows a power-law at even  smaller angular scales $(\ell \le 1300)$, and they found a slightly smaller value  $\beta=1.8$. \citet{Choudhuri2016a,Choudhuri2020} used the TIFR GMRT Sky Survey (TGSS) \citep{Sirothia2014,Intema2016} data to estimate  $C_{\ell}$ of the DGSE for different pointing directions distributed all over the sky. This is the first all-sky measurement of  $C_{\ell}$ at this low frequency of  150 MHz. They also found the slopes $\beta$ in the range $2$ to $3$ for two slightly off-galactic pointing directions.

In this paper, we consider the MWA drift scan observations \citep{Patwa2021}, which were originally carried out to measure the EoR 21-cm signal. This observation is carried out at the fixed declination (DEC; $\delta$) of $-26.7^{\circ}$, and it covers a region of the sky from right ascension (RA; $\alpha$) $349^{\circ}$ to $70.3^{\circ}$ at an interval of $0.5^{\circ}$ along $\alpha$. Visibilities are dumped every 2 min with a total of 163 different pointings centers (PCs). This drift scan observation covers both EoR~0~($0^{\circ}, -26.7^{\circ}$) and EoR~1~($60^{\circ}, -26.7^{\circ}$), which are two of the main targets of MWA EoR 21-cm experiments \citep{Beardsley2016}. It is particularly important to quantify the foregrounds in this region of the sky where a substantial effort is underway to detect the EoR 21-cm signal. Although this observation has 163 PCs, we have analyzed a total of 24 PCs in this paper. Out of the 24 PCs, 21 PCs span the $\alpha$ range from $349^{\circ}$ to $70.3^{\circ}$ at regular $4^{\circ}$ intervals, and 3 PCs are located at intermediate RAs. The interval of $4^{\circ}$ along the RA is sufficient to capture the variation of intensity of the DGSE.  We have used the Tapered Gridded Estimator (TGE; \citealt{Choudhuri2016b}) to characterize the DGSE for this observation. The full-width-at-half-maximum (FWHM) of the MWA primary beam is $27^{\hspace{0.5mm}\circ}$ at $154.2 \, {\rm MHz}$ \citep{Franzen2016}, and we used a window with FWHM $15^{\circ}$ to taper the response of the primary beam. We are able to fit the data with a model only for 6 PCs at $\alpha = 352.5^{\circ}, 353^{\circ}, 357^{\circ}, 4.5^{\circ}, 4^{\circ}$ and $1^{\circ}$. A brief outline of this paper follows: In Section~\ref{sec:method}, we discuss the data analysis and methodology. The results of this study are discussed in Section~\ref{sec:results}, and we summarise and conclude in Section~\ref{sec:sum}.

\section{Methodology}
\label{sec:method}
In this section, we describe the methodology used to estimate the angular power spectrum $C_{\ell}$ of the DGSE using MWA drift scan observations. Here, we have used the Phase II compact configuration of the MWA radio telescope (\citealt{Lonsdale2009,Wayth2018}). The maximum extent of this configuration is $488 {\rm m}$, which is suitable for studying large-scale diffuse emission from the sky. We consider a particular drift-scan observation (project ID G0031; \citealt{Patwa2021}) with an observing time of 5~hr 24~min per night, and the same sky is observed for 10 consecutive nights. Since the observations cover the same region of the sky everyday, we perform Local Sidereal Time (LST) stacking of the measured data (\eg \citealt{Bandura2014, Amiri2022}) and obtain the equivalent one-night drift scan data. The $\delta$ for this drift scan observation is fixed at $-26.7^{\circ}$, and $\alpha$ changes from $349^{\circ}$ to $70.3^{\circ}$. Figure 1 of   \citet{Chatterjee2024} shows the total sky coverage for this observation. The bandwidth of this observation is $30.72 \, {\rm MHz}$, centered at $\nu_c=154.2 \, {\rm MHz}$ and the total bandwidth is divided into 24 coarse-bands with 32 channels each. Frequency resolution of each channel is $\dnu_c=40 \, {\rm kHz}$.

The flagging and calibration details for these data sets are presented in \citet{Patwa2021}. Here, we have used the calibrated visibilities to make a continuum image of angular extent $30^{\circ} \times 30^{\circ}$ centred on our PC. We have used the multi-scale CLEAN feature of \textsc{WSClean} \citep{Offringa2014, Offringa2017} with a cleaning threshold of $3\sigma$ and `Briggs -0.1' weighting-scheme. We have used only the longer baselines ($|u| > 50\lambda$) during imaging in order to avoid large-scale diffuse emission during the deconvolution process. This step will help to model the bright sources only, and we plan to remove them from the total visibility data to study the residual large-scale diffuse emission. The resolution of the final images are relatively poor, as an example for $\alpha = 4.5^{\circ}$ the FWHM of synthesize beam is $\approx 18.5^{'} \times 11.0^{'}$ with a position angle of $42.3^{\circ}$. We have identified and modelled sources with flux density  $S > S_{c} \approx 3 \sigma \, (= 430 \, {\rm mJy})$ from the entire image, where $\sigma$ is the r.m.s. noise estimated from a source free region (see Table~\ref{tab:table1}). We have subtracted out the model visibilities corresponding to the CLEAN component of our sources from the visibility data and used this for the subsequent analysis. To validate our source identification,  we have used \texttt{PyBDSF} \citep{Mohan2015} to extract a source catalogue from our primary beam-corrected image. For source identification, we use a central region of radius $7.5^{\circ}$ where the primary beam is quite well quantified. We have compared the angular position and flux of our sources with those in the GLEAM survey \citep{Wayth2015}. We found the maximum deviation are less than 50 arcsec for all source, whereas the median flux deviation remain less than 25 percent for the sources above $800 {\rm mJy}$. We show the comparison results in ~\ref{appen1} for one pointing only centerd at $(\alpha,\delta=4.3^{\circ},-26.7^{\circ})$ for validation and found that the source properties match quite reasonably with the GLEAM catalogue.


In this paper, we have used a visibility-based power spectrum estimator, namely the TGE, to estimate  $C_{\ell}$ for the data both before and after source removal. The detailed mathematical formalism of the TGE has been discussed in several earlier works \citep{Choudhuri2016b, Choudhuri2017a, Choudhuri2020}, and we briefly summarise the main features here. First, TGE uses gridded visibilities to reduce the computation.  Second, the TGE  tapers the sky response with a tapering function ${\cal W}(\theta)$ that suppresses the contribution from the outer region of the primary beam. Here, we have used a Gaussian ${\cal W}(\theta)=e^{-\theta^2/\theta^2_w}$ that peaks around $\vtht=0$ and falls off rapidly away from the centre. The tapering function ${\cal W}(\theta)$ has a FWHM $\theta_{\rm FWHM}=\theta_w/0.6$ and for this work we have used $\theta_{\rm FWHM}= 15^{\circ}$. Our earlier study \citep{Chatterjee2022} shows that the choice of the FWHM of $15^{\circ}$ results in reasonable SNR values while keeping the foreground contamination from far field sources to a minimum. The tapering is implemented by convolving the measured visibilities with $\tilde{w}(\mathbf{u})$, the Fourier transform of ${\cal W}(\theta)$. Considering a square grid in the $uv$ plane, $\V_{cg}$ the convolved visibility at a grid point  $g$ can be written as

\begin{equation}
    \V_{cg} = \sum_{i}\tilde{w}(\mathbf{u}_g-\mathbf{u}_i) \, \V_i
\label{eq:a2}
\end{equation}

where $\mathbf{u}_g$ refers to the baseline corresponding to the  grid point $g$,  and $\V_i$ is the visibility measured at baseline $\mathbf{u}_i$. Third, the TGE provides an unbiased estimate of the true sky signal by subtracting the noise bias that arises due to the self-correlation of the measured visibilities. 
The TGE is given by 

\begin{equation}
{\hat E}_g= M_g^{-1} \, \left( \left| \V_{cg} \right|^2 -\sum_i \left|
\tilde{w}(\mathbf{u}_g-\mathbf{u}_i) \right|^2 \left| \V_i \right|^2 \right) \,,
\label{eq:a6}
\end{equation}

where $\langle {\hat E}_g \rangle =C_{\ell_g}$ with $\ell_g= 2 \pi \left| \mathbf{u}_g \right|$.  Here, $M_g$ is a normalising factor, which we calculated using simulated visibilities corresponding to a unit angular power spectrum (UAPS), $C_{\ell}=1$. To reduce the statistical fluctuations, we have used $50$ realisations of UAPS to estimate $M_g$.  \citet{Chatterjee2022} presents an extensive description of the simulations used to estimate $M_g$ for MWA observations. It has further validated the TGE, considering simulated MWA observations. 

We have divided the $uv$ plane into annular rings and binned the estimated $C_{\ell_g}$, assuming the sky signal to be statistically isotropic in the plane of the sky. We finally have estimates of the binned $C_{\ell}$ at the mean $\ell$ value corresponding to each bin. To avoid the effect of bandwidth smearing, we have used 17 channels of total bandwidth 0.68 MHz centred at 154 MHz for further analysis. We further averaged those 17 channels to make an equivalent single-channel data for $C_{\ell}$ estimation, presented in the next section.

\section{Results}
\label{sec:results}

\begin{figure*}[t]
    \centering
    \includegraphics[width = \textwidth]{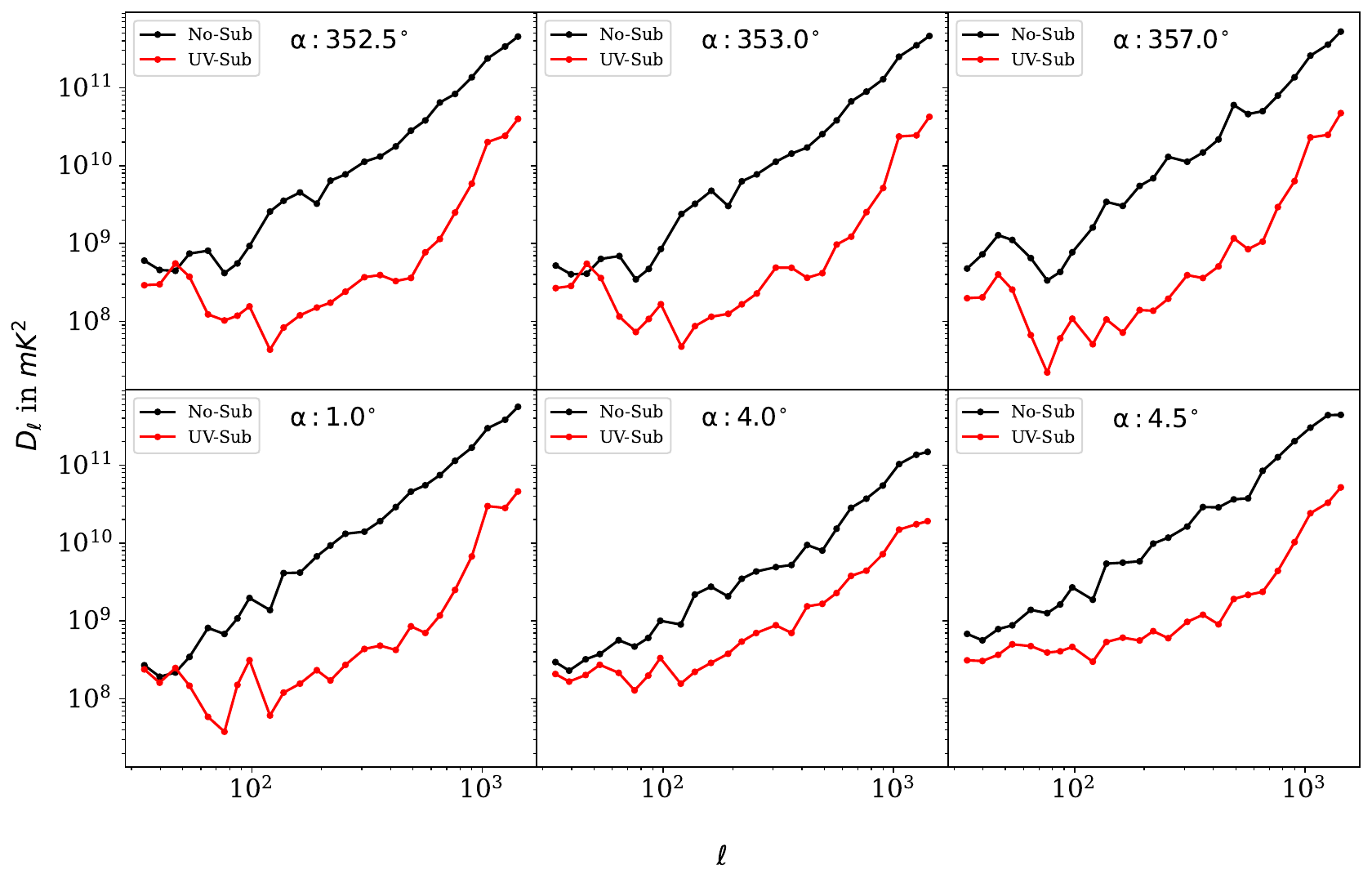}
    \caption{This shows the estimated ${\cal D}_{\ell}=\ell(\ell+1)C_{\ell}/2\pi$ as a function of $\ell$ for six pointing centered at  
$\alpha=352.5^{\circ}, 353^{\circ}, 357^{\circ}, 4.5^{\circ}, 4^{\circ}$ and $1^{\circ}$, and $\delta$ remain the same for all pointing at 
$\delta=-26.7^{\circ}$. The black lines show the total data before point source subtraction, and the red lines show the $C_{\ell}$ after removing the sources above $3\sigma$.}
    \label{fig:cl_nf}
\end{figure*}

Figure \ref{fig:cl_nf} shows the measured mean-squared brightness temperature fluctuation 
\begin{equation}
  {\cal D}_{\ell}=\ell(\ell+1){ \, \cal C}_{\ell}/ \, 2\pi  
\label{eq:dl}
\end{equation}
for both before (No-Sub) and after (UV-Sub) point source subtraction from the 6 PCs where we are able to measure the contribution of the DGSE from the residual visibility data. In each panel, the red and black lines show the `No-Sub' and `UV-Sub' cases, respectively. We find that, in the No-Sub scenarios, $D_{\ell}$ ranges from approximately $\sim 2 \times 10^{8}$ mK$^{2}$ to $\sim 5 \times 10^{11}$ mK$^{2}$ for $\ell$ values in the range $40$ and $1000$. The amplitude of the ${\cal D}_{\ell}$ is consistent with earlier observations with the GMRT at the same frequency range \citep{Choudhuri2017a,Choudhuri2020}. We also note that ${\cal D}_{\ell} \propto (\ell/1000)^2$ for $\ell > 200$ for all pointings, which indicates that the measured ${\cal D}_{\ell}$ is dominated by the Poisson fluctuations due to bright point sources \citep{Ali2008}. Next, to study the statistical properties of the DGSE, we have removed the point sources with flux  $S > S_c \approx 3\sigma$ (Table~\ref{tab:table1})  from the data and used the residual data to estimate ${\cal D}_{\ell}$. After removing the bright sources, we see that the amplitude falls significantly across the whole $\ell$ range. However, we clearly see two distinct $\ell$ ranges in the residual  ${\cal D}_{\ell}$. After point source subtraction, the amplitude decreases by a larger amount at the high $\ell$ values $(\ell >  200)$ compared to the smaller $\ell$. At high $\ell$, we find again that ${\cal D}_{\ell} \propto (\ell/1000)^2$. The sky signal in this $\ell$ range is mainly dominated by the Poisson fluctuations from the point sources that are below our flux limit $S \le S_c $, which we are not able to subtract out. We expect the sky signal ( ${\cal D}_{\ell}$ ) at this  $\ell$ range to go down even further if we have more sensitive observations where we can achieve a lower value of $S_c$. At lower $\ell$ ($ \le 200$),  the sky signal does not go down as much after point source subtraction. The slope here is shallower than ${\cal D}_{\ell} \propto \ell^2$. We believe that the DGSE starts to dominate the sky signal at $\ell \le 200$ after point source subtraction. Further, we do not expect the sky signal at the DGSE-dominated small $\ell$ $( \le 200)$ range to go down much, even if it is possible to lower $S_c$ and improve point source subtraction.

\begin{figure*}[t]
    \centering
    \includegraphics[width = \textwidth]{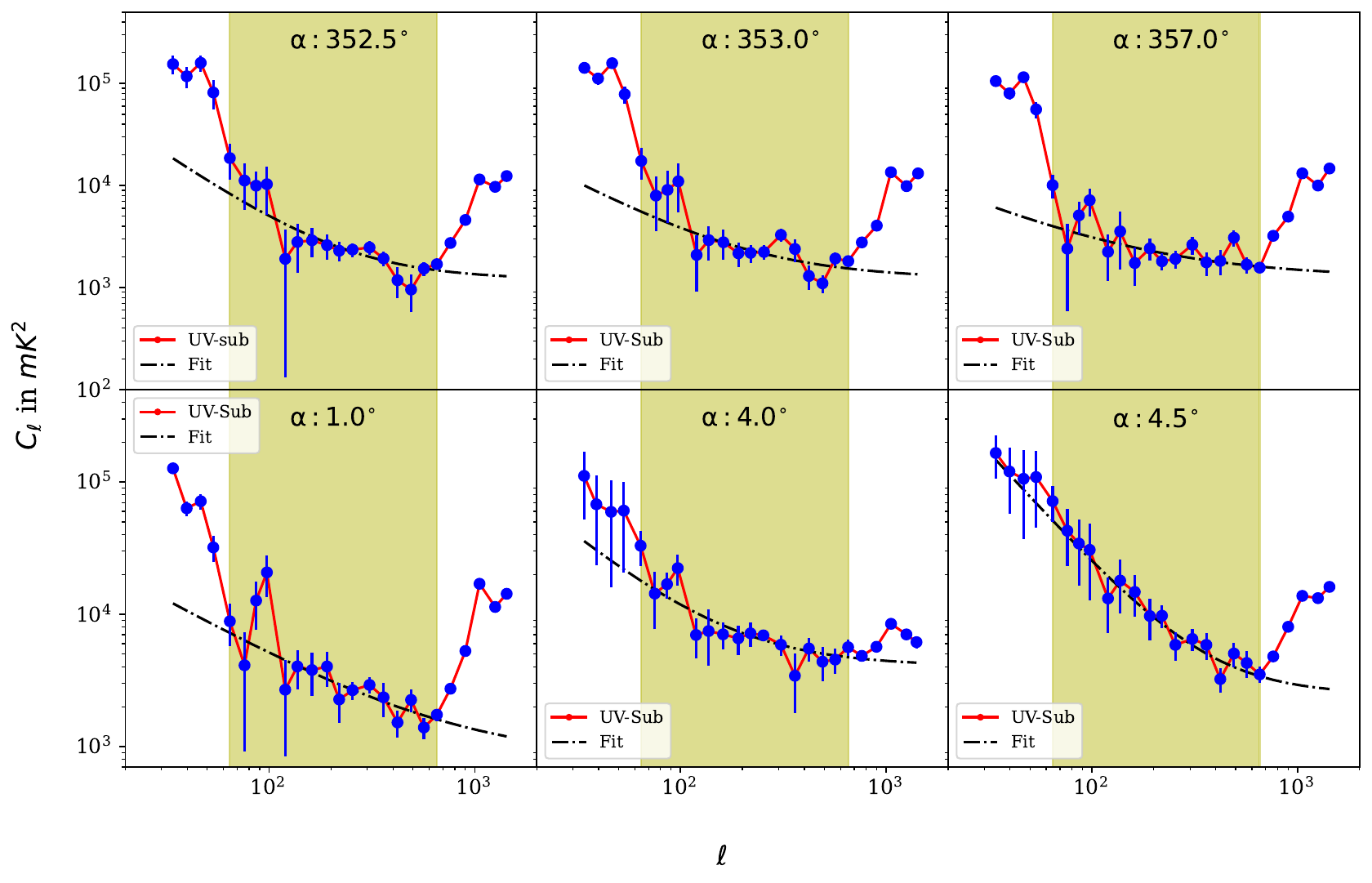}
    \caption{The blue points show the estimated angular power spectrum ${\cal C}_{\ell}$ as a function of $\ell$ with $1\sigma$ error bars from the residual data. The black dot-dashed line shows the model ${\cal C}^{\rm M}_{\ell}$ (eq.~\ref{eq:fit}) with best-fitted parameters from the MCMC run. The shaded region $(65 \le \ell \le 650)$ shows the data range used for the fitting.} 
    \label{fig:cl_f}
\end{figure*}

Figure \ref{fig:cl_f} shows the estimated ${\cal C}_{\ell}$ (red solid line) of the residual sky signal for the UV-Sub case. Here, we show the results for aforementioned 6 PCs. We interpret the measured ${\cal C}_{\ell}$ as a combination of a power law $(\propto \ell^{-\beta})$ due to DGSE and a constant  Poisson fluctuation part due to the residual point sources. Here, we have used 
\begin{equation}
{\cal C}^M_{\ell}=A\times \bigg(\frac{1000}{\ell}\bigg)^{\beta}+C
\label{eq:fit}
\end{equation}
to model the residual ${\cal C}_{\ell}$. Later, we used the Markov Chain Monte Carlo (MCMC) ensemble sampler to estimate the best-fit values and errors for the model parameters $A,\beta$ and $C$. To estimate the measurement errors, we assume that the residual sky signal is a realisation of a Gaussian random field with a given angular power spectrum. In reality,  it is quite likely that this assumption is not strictly valid for the actual data. However, little is known about the statistics of the residual signal, and this assumption considerably simplifies the analysis.  To estimate the $1 \sigma$ error bars of the estimated  ${\cal C}_{\ell}$, we have simulated $40$ independent Gaussian random realisations of the residual sky signal and estimated the resulting visibilities. The sky signal was simulated so that we roughly recover the estimated ${\cal C}_{\ell}$  if we analyse the simulated visibilities in exactly the same way as the actual data. The simulated visibilities also include a Gaussian random system noise contribution with the same r.m.s. in the actual data. In this observation, we expect the r.m.s. noise is likely dominated by thermal receiver noise, and the confusion noise from unresolved point sources is insignificant for the pointing considered here. We have used the r.m.s. of the $40$ independent relizations of the simulated ${\cal C}_{\ell}$ to estimate the $1 \sigma$ errors for the estimated ${\cal C}_{\ell}$. 

We expect ${\cal C}_{\ell}$ measured at low $\ell$ to be affected by the convolution with the primary beam and the tapering function. Using realistic simulations, \citet{Chatterjee2022} showed that the range $  \ell \ge 65$ is largely unaffected by convolution for the tapering window function, and we have excluded the range $\ell < 65$ for fitting eq.~(\ref{eq:fit}). We further notice that the residual ${\cal C}_{\ell}$ increases sharply at $\ell > 650$.  Our data has poor sampling at the longer baselines, and it is possible that the source subtraction is not very effective at these small angular scales. Here, we have also discarded the range $\ell > 650$ for fitting eq.~(\ref{eq:fit}).

We use the $\ell$ range $65 < \ell < 650$ to fit the model ${\cal C}^M_{\ell}$ (eq.~\ref{eq:fit}) with the measured values from the data. As mentioned, we have used the MCMC ensemble sampler to get the best-fit values of the model parameters $A,\beta$ and $C$. Here, we have used the python module \texttt{ EMCEE} \citep{emcee}, which implements the affine-invariant ensemble sampling (\citealt{ESAI}) algorithm, to get the posterior probability distribution of the parameters. We assumed a Gaussian likelihood function for this analysis. Also, we set a uniform prior of the parameters in the ranges: 
$\mathcal{U}(0,\infty)$ on $A$, $\mathcal{U}(0,5.0)$ on $\beta$ and $\mathcal{U}(0, \infty)$ on $C$. We put the condition that $A$ would always be positive based on the earlier analysis of the DGSE \citep{Choudhuri2017a}. In the MCMC run, we used 32 random walkers initialised randomly, and we ran the full chain for 100000 steps, out of which 10000 steps are discarded for the burn-in process. We show the best-fit model ${\cal C}^M_{\ell}$ after the MCMC run, along with the measured ${\cal C}_{\ell}$ from the MWA observations. The black dot-dashed lines in Figure \ref{fig:cl_f} show the best-fit model ${\cal C}_{\ell}^M$ (eq.~\ref{eq:fit}) after MCMC run for these 6 PCs. This figure also highlights the range  $65 \le \ell \le 650$ that has been used for model fitting. We see that the model fits quite well the measured ${\cal C}_{\ell}$ in this range, and the reduced chi-square $(\chi_{\rm R}^2)$ values for the fit, and the $p$-value \footnote{The probability-to-exceed (PTE or p-value)  is the probability of obtaining a value of $\chi^{2}$  higher than what is obtained, and defined as $p = 1 - \text{CDF}(\chi^2; f)$ where CDF is the cumulative distribution function of the $\chi^{2}$ distribution and $f$ is the number of degrees of freedom. A low p-value would imply that there is an unlikely chance of obtaining a higher $\chi^{2}$ than what is already obtained. It suggests that the model and the data are more distinct from each other than what a random chance would allow (see e.g. \citealt{rice2006msda}). } are given in Table~\ref{tab:table1}. In the figure, we also plot the ${\cal C}^M_{\ell}$ beyond the range $65 \le \ell \le 650$, used for the fitting. This is to show the expected sky contribution beyond this fitted region without any noise and convolution error.

\begin{table*}[h!]
\renewcommand{\arraystretch}{1.5} 
\setlength{\tabcolsep}{10pt}      
\begin{center}
\resizebox{\textwidth}{!}{       
\begin{tabular}{c c c c c c c c}
\hline
$\alpha$ & $\sigma$ (in mJy) & $A$ (in mK$^{2}$)& $\beta$ & $C$ (in mK$^{2}$)& $\chi^{2}_{R}$ & p-value  \\
\hline
 $352.5^{\circ}$ & $146.4$ & $154.3^{\hspace{0.5mm}+397.5}_{\hspace{0.5mm}-114.6}$ & $1.4^{\hspace{0.5mm}+0.6}_{\hspace{0.5mm}-0.6}$ & $1199.1^{\hspace{0.5mm}+292.5}_{\hspace{0.5mm}-540.2}$ & $1.147$ & $0.313$  \\
\hline
$353.0^{\circ}$ & $150.0$ & $201.9^{\hspace{0.5mm}+525.3}_{\hspace{0.5mm}-167.4}$ & $1.1^{\hspace{0.5mm}+0.8}_{\hspace{0.5mm}-0.5}$ & $1214.6^{\hspace{0.5mm}+337.7}_{\hspace{0.5mm}-633.5}$ & $2.185$ & $0.008$ \\
\hline
$357.0^{\circ}$ &$143.5$& $241.2^{\hspace{0.5mm}+603.2}_{\hspace{0.5mm}-200.3}$ & $0.9^{\hspace{0.5mm}+0.8}_{\hspace{0.5mm}-0.5}$ & $1255.9^{\hspace{0.5mm}+348.9}_{\hspace{0.5mm}-674.5}$ & $1.585$ & $0.081$  \\
\hline
$1.0^{\circ}$ & $177.9$& $568.4^{\hspace{0.5mm}+398.9}_{\hspace{0.5mm}-314.9}$ & $0.9^{\hspace{0.5mm}+0.3}_{\hspace{0.5mm}-0.2}$ & $777.7^{\hspace{0.5mm}+471.8}_{\hspace{0.5mm}-494.7}$ & $1.234$ & $0.247$ \\
\hline
$4.0^{\circ}$ &$146.0$& $388.3^{\hspace{0.5mm}+1133.5}_{\hspace{0.5mm}-297.9}$ & $1.3^{\hspace{0.5mm}+0.6}_{\hspace{0.5mm}-0.6}$ & $4057.5^{\hspace{0.5mm}+916.9}_{\hspace{0.5mm}-1675.1}$ & $1.018$ & $0.430$  \\
\hline
$4.5^{\circ}$ & $143.3$& $407.7^{\hspace{0.5mm}+353.7}_{\hspace{0.5mm}-211.0}$ & $1.7^{\hspace{0.5mm}+0.3}_{\hspace{0.5mm}-0.3}$ & $2574.8^{\hspace{0.5mm}+679.0}_{\hspace{0.5mm}-777.0}$ & $0.656$ & $0.839$ \\
\hline
\end{tabular}
}
\caption{This table provides the details of the model fitting for 6 PCs. The column descriptions are as follows: (1) RA of the pointings, (2) rms of the image, (3) (4) (5) the best-fit value of parameter $A$, $\beta$ and $C$ after MCMC run ( eq.~\ref{eq:fit}), (6) $\chi^{2}_{R}$, and  (7) p-value.}
\label{tab:table1}
\end{center}
\end{table*}

\begin{figure}[t]
    \centering
    \includegraphics[width =\textwidth]{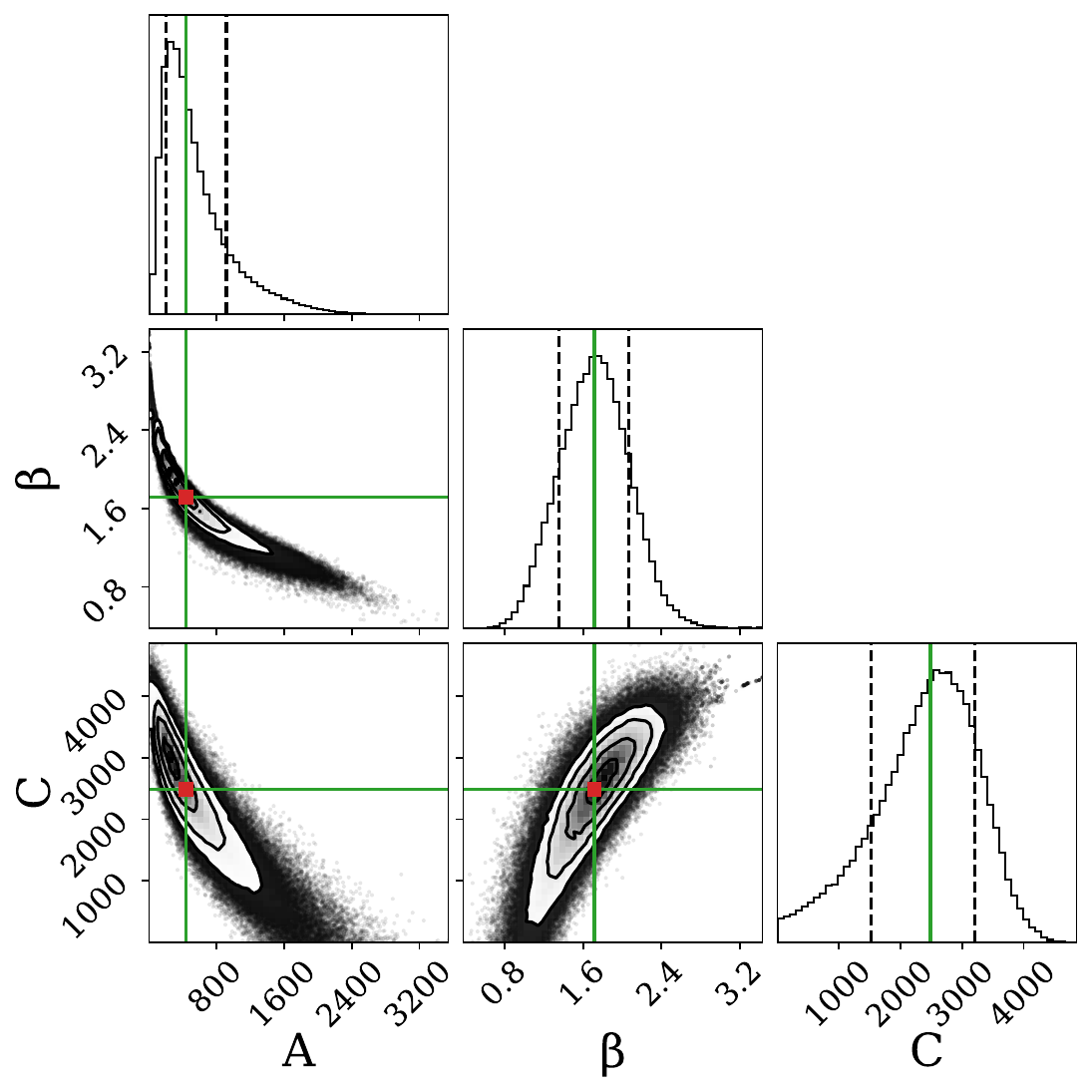}
    \caption{The posterior distributions of the parameters A, $\beta$ and C for PC centered at $\alpha=4.5^{\circ}$. The best-fit values of the parameters and their corresponding errors are as follows: $A=407.7^{\hspace{0.5mm}+353.7. }_{\hspace{0.5mm}-211.0}$, $\beta=1.7^{\hspace{0.5mm}+0.3}_{\hspace{0.5mm}-0.3}$ and $C=2574.8^{\hspace{0.5mm}+679.0}_{\hspace{0.5mm}-777.0}$. We see that the parameters $\beta$ and C are positively correlated, while A shows a negative correlation with both.} 
    \label{fig:corner-3720}
\end{figure}

Figure \ref{fig:corner-3720} shows the posterior distributions of the parameters $A$, $\beta$ and $C$ for one PC centered at $\alpha=4.5^{\circ}$. The diagonal panels show the one-dimensional marginalised posterior distributions for the parameters. The off-diagonal panels show the 2D contours of correlations between parameter pairs. The vertical lines from the left are the 16th (black), 50th (green) and the 84th (black) percentile, respectively. The best-fit values of the parameters and their corresponding errors are as follows: $A=407.7^{\hspace{0.5mm}+353.7}_{\hspace{0.5mm}-211.0}$, $\beta=1.7^{\hspace{0.5mm}+0.3}_{\hspace{0.5mm}-0.3}$ and $C=2574.8^{\hspace{0.5mm}+679.0}_{\hspace{0.5mm}-777.0}$. We see that all the parameters are strongly correlated. The parameters $\beta$ and C are positively correlated, while $A$ shows an anti-correlation with both. The Pearson’s product-moment correlation coefficients for $A$  with $\beta$ and $C$ are -0.9 and -0.89, respectively, whereas the same for $\beta$ with $C$ is 0.84. We have shown the posterior distribution of rest of the PCs in \ref{appen2}. 

In Table~\ref{tab:table1}, we report the median value, which represents the 50$^{\mathrm{th}}$ percentile, alongside the upper and lower limits corresponding to the 84$^{\mathrm{th}}$ and 16$^{\mathrm{th}}$ percentiles, respectively. We find that the amplitude $A$ varies significantly for different pointings. The pointing centered at $\alpha=352.5^{\circ}$ has the lowest amplitude $154.3 {\rm mK^2}$, where as the 
$\alpha=1^{\circ}$ has the highest amplitude of $568.4 {\rm mK^2}$. We expect the residual $C_{\ell}$ to be dominated most likely by the DGSE, and the amplitude may vary depending on the location of the PC with respect to the Galactic plane. We have seen a similar variation of the amplitude at different pointing centers using the all-sky TGSS survey \citep{Choudhuri2020}. The values of $\beta$ vary in the range $0.9$ to $1.7$, which are consistent with other measurements in this frequency range \citep{Bernardi2009,Ghosh2012, Iacobelli2013a,Choudhuri2016a}. The constant part $C$ is coming due to unsubtracted point sources in the field and varies in the range $777$ to $4057$ mK$^{2}$ for different pointing. To show the goodness of the fit, we quoted the value of $\chi^{2}_{R}$ and p-values for all pointings in Table~\ref{tab:table1}. We see that 5 out of the 6 PCs have the $p$-values $>0.05$, which implies that the model provides a reasonable fit to the data for these PCs. Considering the PC centered at $\alpha=353^{\circ}$, we find the $p$-value to be $0.008$. Although the fit is poorer for this PC, there is still a $0.8\%$ chance for the model to fit the data, and we have considered this to be acceptable.

\begin{figure}
    \hspace{-1cm}
    \includegraphics[width=\textwidth]{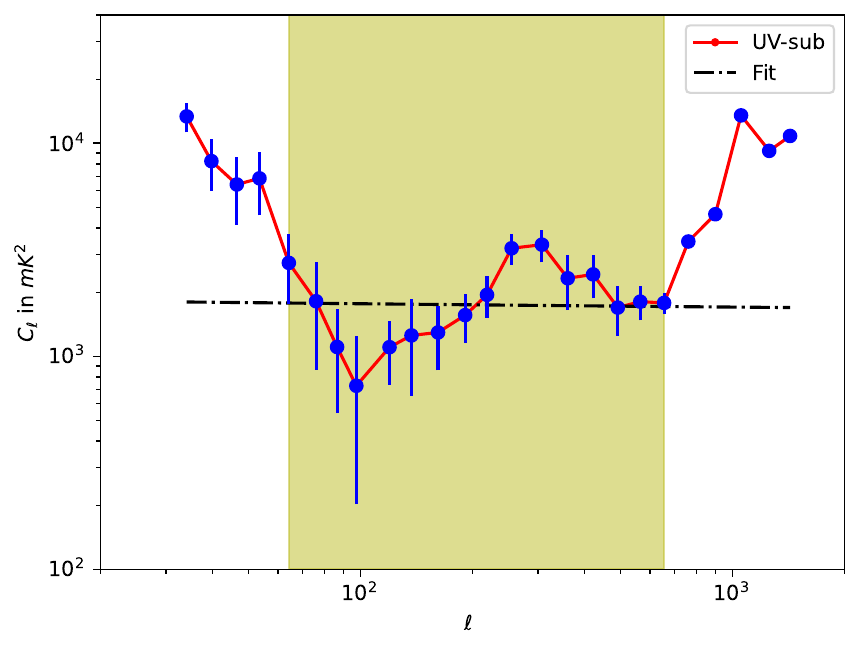}
    \caption{The blue points show the estimated ${\cal C}_{\ell}$ as a function of $\ell$ with $1-\sigma$ error bars from the residual data for a PC centered at $\alpha=37^{\circ}$. The black dot-dashed line shows the fitted model, which is a straight line with amplitude $A=272.8^{\hspace{0.5mm}+719.2}_{\hspace{0.5mm}-238.6}$. Here, we are not able to fit the data points with a power-law model (eq.~\ref{eq:fit}).}
    \label{fig:discardedpc}
\end{figure}

We have identified 6 PCs (out of the 24 PCs) for which the residual angular power spectra $C_{\ell}$ could be fitted with a model $C_{\ell}^M$ given in eq.~\ref{eq:fit}. For the remaining PCs, the measured $C_{\ell}$ does not exhibit a power-law behavior, and we show one such representative PC $(\alpha=37^{\circ})$ in Figure~\ref{fig:discardedpc}. Here, we see that the best-fit model $C_{\ell}$ (black dot-dashed) is nearly flat across the entire $\ell$ range. We attribute this flatness to the unclustered Poisson distribution of faint point sources within this PC \citep{Ali2008}. These sources have flux densities below the $3\sigma$ threshold and thus could not be subtracted using the standard technique. Further, imaging artifacts around bright sources, possibly arising from residual gain calibration errors, may affect the measured $C_{\ell}$. We note that the observations that we consider here are shallow ($\sim 17$ minutes per PC), and the limited baseline coverage results in poor angular resolution, and these make it difficult to subtract compact sources accurately, which in turn leads to the failure of fitting a model to the $C_{\ell}$. We also note that the bright source Fornax A starts to dominate the visibilities during the latter part of the observation $\alpha \gtrsim 40^{\circ}$ \citep{Chatterjee2024}, which makes it difficult to model the residual visibilities in that region.

\section{Summary and Conclusions}
\label{sec:sum}
In this paper, we studied the statistical properties of the DGSE in terms of the angular power spectrum. For this purpose, we used the drift scan observations of the Phase II compact configuration of the MWA. Here, we analyze total
24 PCs to characterize the DGSE, however, we are only able to fit the data with a model for 6 PCs centered at $\alpha=352.5^{\circ}, 353^{\circ}, 357^{\circ}$,\\$4.5^{\circ}, 4^{\circ}$ and $1^{\circ}$. The total observing time for each pointings is of around ~17~mins (a total 10 nights of LST stacking). We removed all the bright point sources above $3\sigma$ (Table~\ref{tab:table1}) to detect the DGSE, which is the most dominant component at low-frequency observation after point source removal.

We apply the TGE to measure the ${\cal D}_{\ell}$ (eq.~\ref{eq:dl}) of the DGSE. In Figure \ref{fig:cl_nf}, we show both the measured ${\cal D}_{\ell}$ before and after the point source removal. The value of the ${\cal D}_{\ell}$ before point source subtraction varies from $\sim 2 \times 10^{8}$ mK$^{2}$ to $\sim 5 \times 10^{11}$ mK$^{2}$ for $\ell$ values in the range $40$ and $1000$. Also, it follows a power law ${\cal D}_{\ell} \propto \ell^2$ for $\ell >  200$. This behaviour of ${\cal D}_{\ell}$ is due to the Poisson fluctuations of the bright point sources, and it is consistent with the model prediction \citep{Ali2008}. We 
removed the bright sources with flux  $S > S_c \approx 3 \sigma$ from the data and measured the ${\cal D}_{\ell}$ from the residual data. Here, we clearly see two distinct $\ell$ ranges in the measured ${\cal D}_{\ell}$. At higher $\ell$ values $(\ell >  200)$ values, it follows the same power law ${\cal D}_{\ell} \propto \ell^2$; however, the amplitude falls substantially as compared with the case where the bright sources are not removed. We expect this higher $\ell$ range in the residual data to be dominated by the Poisson fluctuations of the point sources below  $3 \sigma$ level. At lower $\ell$ ($ \le 200$), the slope is shallower than ${\cal D}_{\ell} \propto \ell^2$, and we believe the DGSE dominates at large angular scales or lower $\ell$ range.

We fit the residual ${\cal C}_{\ell}$ with a model as given in eq.~\ref{eq:fit}. Here, the power-law part is due to the large-scale DGSE, and the constant part is due to the unsubtracted point sources at small angular scales. We found that the convolution of the primary beam affects the lower $\ell$ range $(\ell \le 65)$. Also, the large $\ell$ values $(\ell \ge 650)$ are dominated by the system noise due to the limited number of samples in those bins. We use only the range $(65 \le \ell \le 650)$, as shown with a shaded region in Figure~\ref{fig:cl_f}, to fit the measured ${\cal C}_{\ell}$ with the model. We have used the MCMC ensemble sampler to estimate the best-fit values and errors for the model parameters $A,\beta$, and $C$. In Figure~\ref{fig:corner-3720} (and fig.~\ref{fig:corner1}-\ref{fig:corner3}), we show the posterior distribution of those parameters (diagonal panels) and also the correlation of these parameters (off-diagonal panels). We found a strong anti-correlation of parameter $A$ with $\beta$ and $C$, and a strong correlation between $\beta$ and $C$. The best-fit values (50$^{\rm th}$ percentiles) and their uncertainties (16$^{\rm th}$ and 84$^{\rm th}$ percentiles) of parameters  $A,\beta$, and $C$ are given Table~\ref{tab:table1}. The values of $A$ ranges from $154.3$ to $568.4$ mK$^{2}$. This is because of the different contributions of the Galactic emission in different pointings we considered here. The values of $\beta$ changes from $0.9$ to $1.7$ for different pointings. These $\beta$ values are consistent with $2\sigma$ measurement with the earlier measurement in a similar angular and frequency range \citep{Bernardi2009, Ghosh2012, Iacobelli2013a, Choudhuri2017a}. The constant part $C$ is coming due to unsubtracted point sources in the field and varies in the range $777$ to $4457$ mK$^{2}$ for different pointing.

We studied a large patch of the sky in the southern hemisphere using the MWA drift scan observation to characterize the DGSE. Earlier, we did a similar study for the entire sky using the TIFR-GMRT sky survey at 150 MHz \citep{Choudhuri2016b, Choudhuri2020}. Here, we expect the signal to be dominated by large-scale diffuse emission, and we assume it is a Gaussian random field generated by some statistical random process, e.g., MHD turbulence. We can use this to model the diffuse foreground model and subtract from the data for EoR observation in this region. The amplitude of ${\cal C}_{\ell}$ at $\ell=1000$ varies for different pointings considered in this analysis. This variation is expected because the intensity of the Galactic synchrotron emission highly depends on the fluctuations of electron density and the magnetic field of the sky's position. This study will help us to constrain the electron density and the magnetic field strength in this region. Recently, \citet{Chakraborty2019a} have studied the spectral nature of the $C_{\ell}$, and found that the amplitude vs frequency follows a double power law nature with a break at $405$ MHz. Next, we plan to analyse the a large bandwidth of MWA data to characterize  the spectral nature of the $C_{\ell}$ for different pointings. This will help us study the statistical properties of the DGSE for a large area of the sky and at different frequencies.

\section*{Acknowledgements}
S. Chatterjee acknowledges support from the South African National Research Foundation (Grant No. 84156) and the Inter-University Institute for Data Intensive Astronomy (IDIA). IDIA is a partnership of the University of Cape Town, the University of Pretoria and the University of the Western Cape. S. Chatterjee would also like to thank Dr. Devojyoti Kansabanik, for helpful discussions.We acknowledge the use of the ilifu cloud computing facility – www.ilifu.ac.za. S. Choudhuri would like to thank Dr. Nirupam Roy and Dr. Prasun Dutta for useful discussions. S. Choudhuri would also like to SERB-MATRICS for providing financial support. 

\bibliography{mylist}
\label{lastpage}

\appendix
\counterwithin{figure}{section}
\section{Astrometry}
\label{appen1}
\begin{figure*}
    \hspace{-1cm}
    \includegraphics[scale=0.45]{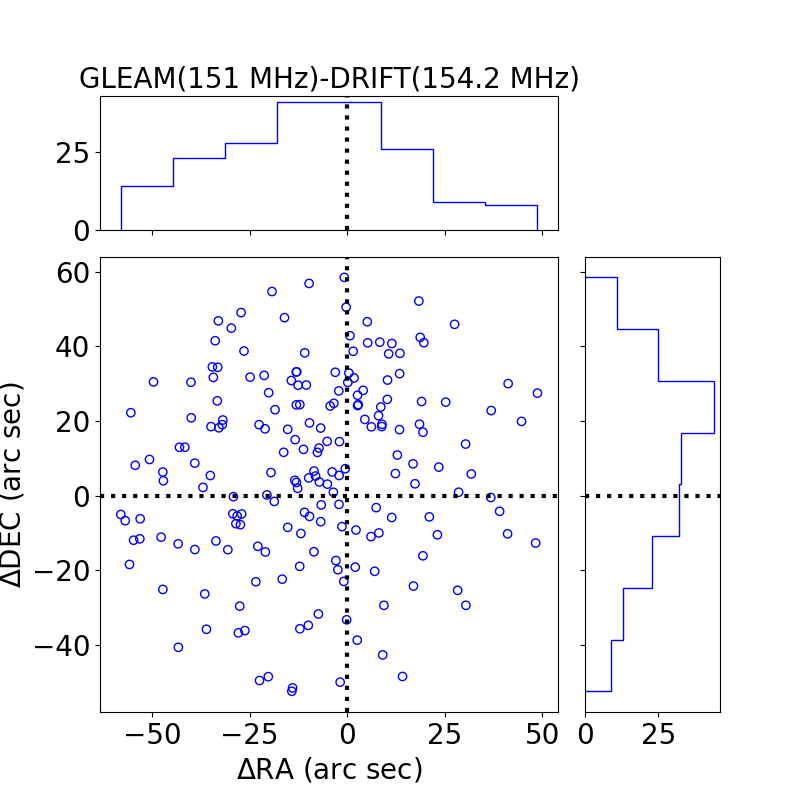}
    \includegraphics[scale=0.45]{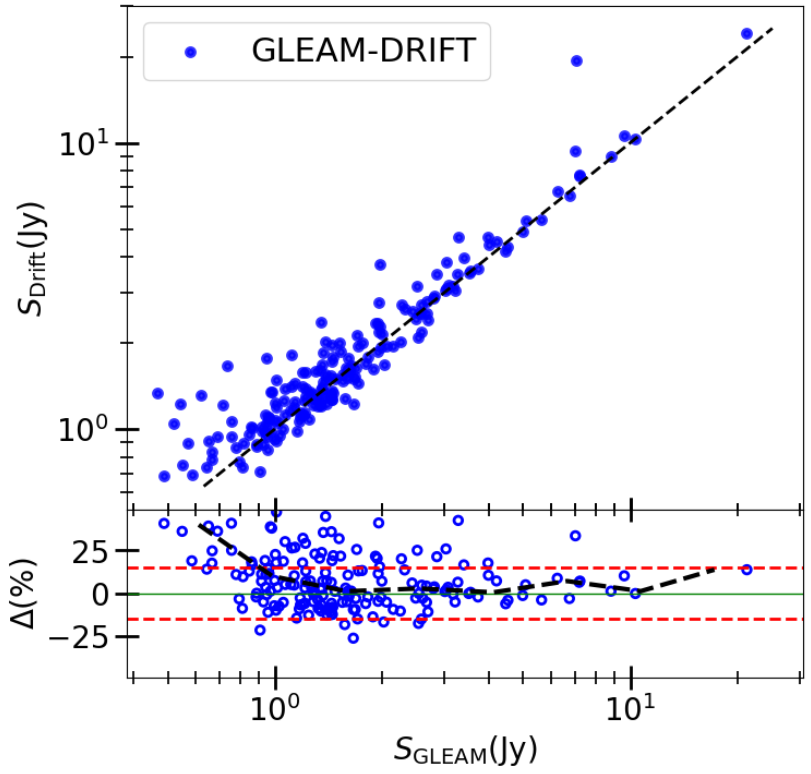}
    \caption{The left panel shows the position offset in terms of $(\Delta RA = \alpha_{\rm Drift} - \alpha_{\rm GLEAM})$ and $(\Delta DEC = \delta_{\rm Drift} - \delta_{\rm GLEAM})$ of the 190 number of sources from this observation and GLEAM catalogue. The upper right panel shows the GLEAM flux values extrapolated from 151 MHz to 154 MHz along the x-axis and the flux recovered from this observation along the y-axis. The lower panel shows the fractional deviation ($\Delta$) of the recovered flux values with respect to the GLEAM catalogue, and the black dashed show the binned median values of the $\Delta$.}
    \label{fig:appen1}
\end{figure*}

In this appendix, we show the comparisons of the point sources extracted using this MWA observation with the GLEAM survey \citep{Wayth2015}. We select all the bright sources above 430 mJy/beam from a region of radius $7.5^{\circ}$ centred at $(\alpha,\delta=4.3^{\circ},-26.7^{\circ})$. The total number of sources we got from this observation is around 400. For rest of the discussion we consider an angular scale of $0.4 \times \theta_{\rm FWHM}$, as the reliability of primary beam modeling plunges after that. This reduces the effective number of sources to 190. First, we compare the position of the sources from these two catalogues. The left panel of Figure \ref{fig:appen1} shows the position offset in terms of $(\Delta RA = \alpha_{\rm Drift} - \alpha_{\rm GLEAM})$ and $(\Delta DEC = \delta_{\rm Drift} - \delta_{\rm GLEAM})$ of the 190 number of sources from these two catalogues. Here, the deviations are less than $50$ arcsec for all sources. Note that the angular resolution of this observation is around $18.54^{'} \times 11.1^{'}$. Also, the deviation is symmetric around zero along RA; however, there is a slight offset $(\sim 30^{''})$ along DEC. The effect of the ionosphere can shift the position of the bright point source, and this can lead to inaccuracies during source subtraction and, potentially, signal loss. In presence of extreme ionospheric activity, performance of the calibration process becomes poor which may introduce systematic offset \citep{Chege2022, Pal2024}. The offset of $(\sim 30^{''})$ along DEC is significantly smaller than the synthesized beam and we expect this to have a very minimal impact on our analysis. We conclude that the recovered position of the sources matches the GLEAM catalogues reasonably. As we are removing these point sources from the data, this offset will not introduce any bias in our final power spectrum measurement of the residual visibility data.

We also compared the recovered flux values with the GLEAM sources for those 190 sources. The upper right panel of Figure \ref{fig:appen1} shows the GLEAM flux values, extrapolated from 151 MHz to 154 MHz using the measured spectral index from the GLEAM survey, along the x-axis and the flux recovered from this observation along the y-axis. The black dashed line shows the $1:1$ line. We see that all the points at higher flux values lie on top of the line; however, for small flux values, the points deviate significantly. The lower panel of the right panel shows the fractional deviation ($\Delta$) of the recovered flux with respect to the GLEAM flux. We see that for most sources, $\Delta$ is less than 25 percent; however, there are few sources for which the $\Delta$ becomes large ($\ge 25$ percent). We expect this flux deviation might be due to the error in the flux calibration for this observation. We are not planning to do any flux correction for this observation, as those sources are removed from the data.

\section{MCMC results for PCs}
\label{appen2}
\begin{figure*}
    \hspace{-1cm}
    \includegraphics[width = 0.5\textwidth]{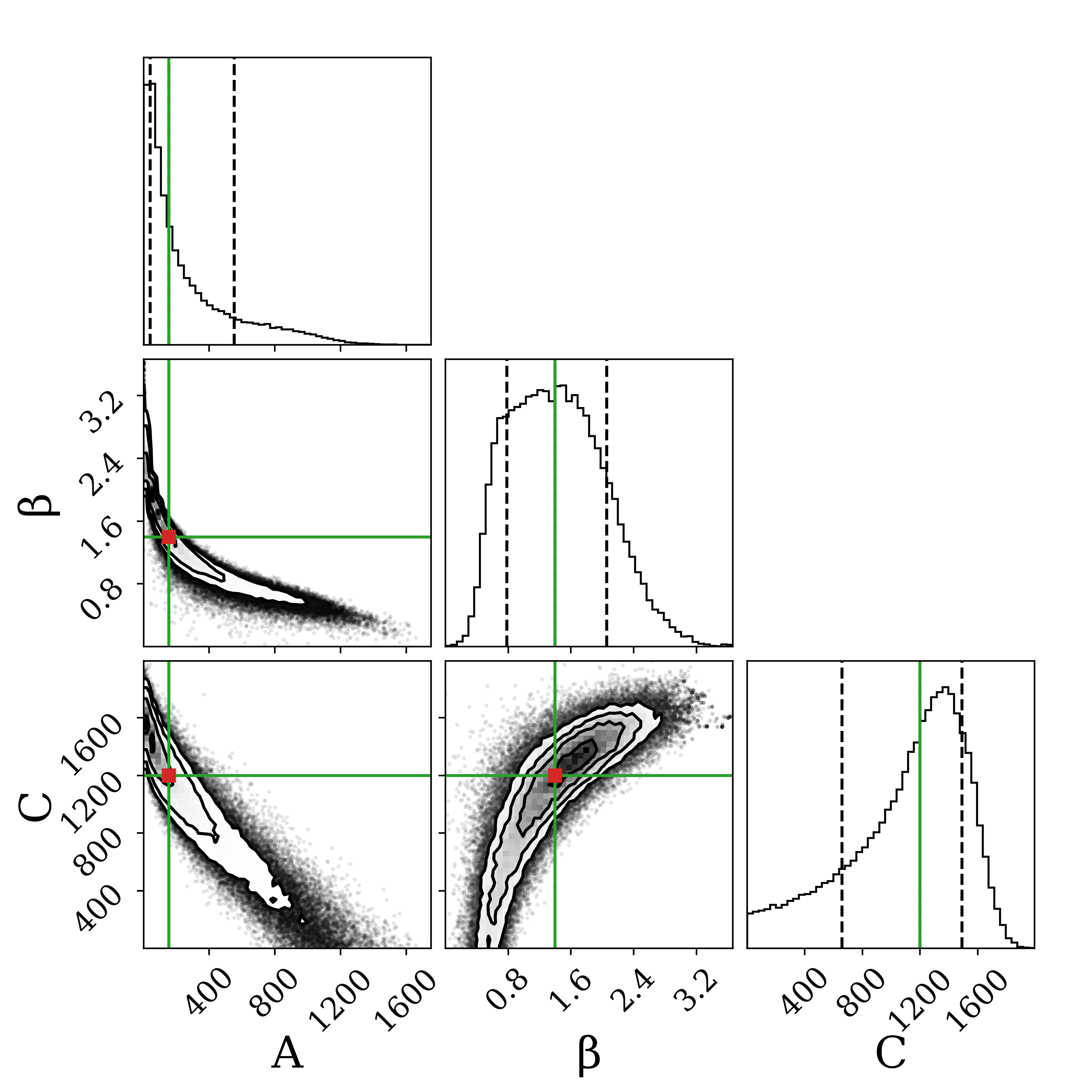}
   \includegraphics[width = 0.5\textwidth]{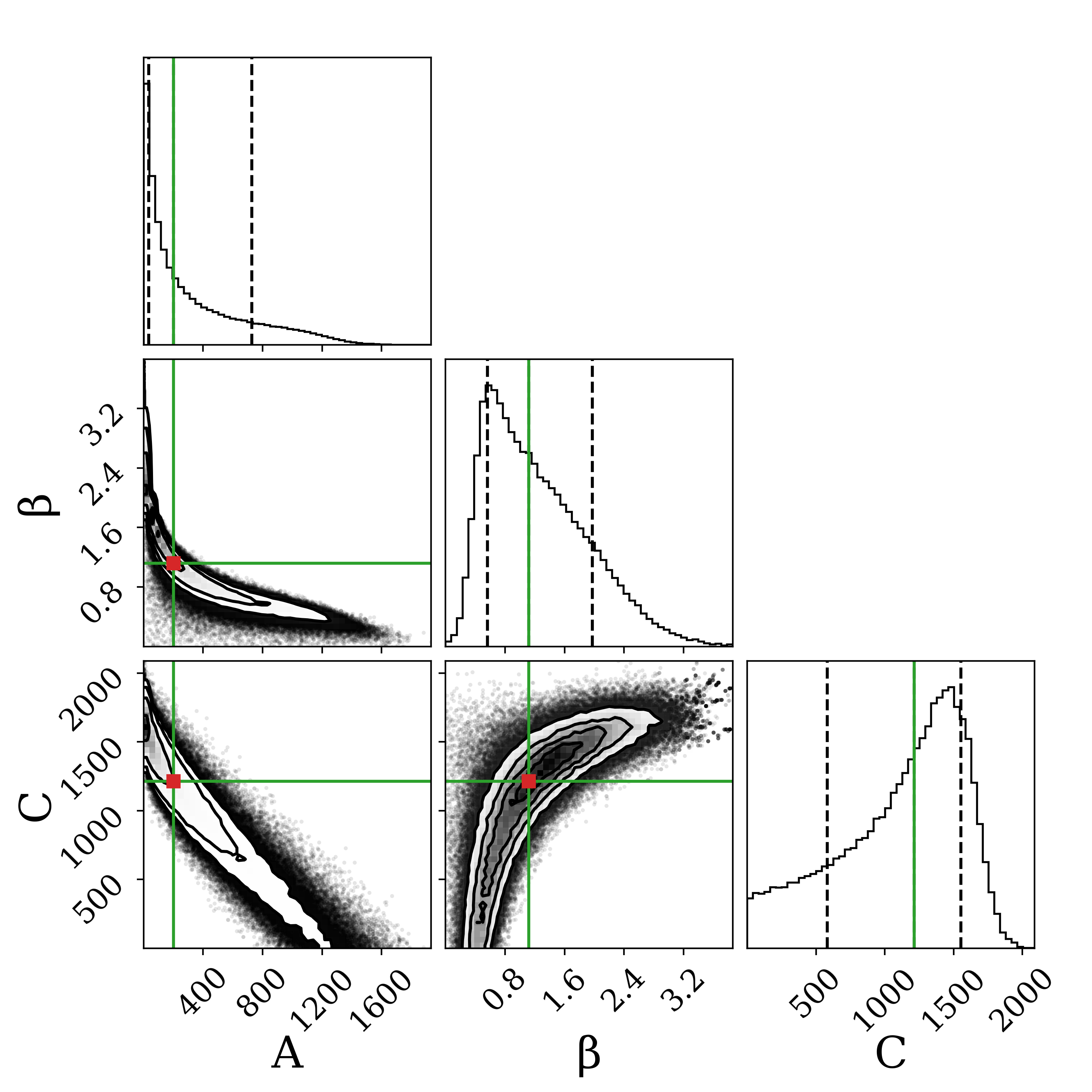}
    \caption{The panels show the posterior probability distribution of parameters $A,\beta$ and $C$ after the MCMC run for $\alpha=352.5^{\circ}$ on the left and $\alpha=353.0^{\circ}$ on the right. The green vertical lines in the diagonal panels show the 50th percentiles of the samples in the marginalised distributions, and two vertical black lines show the 16th and 84th percentiles.}
    \label{fig:corner1}
\end{figure*}

\begin{figure*}
    \hspace{-1cm}
    \includegraphics[width = 0.5\textwidth]{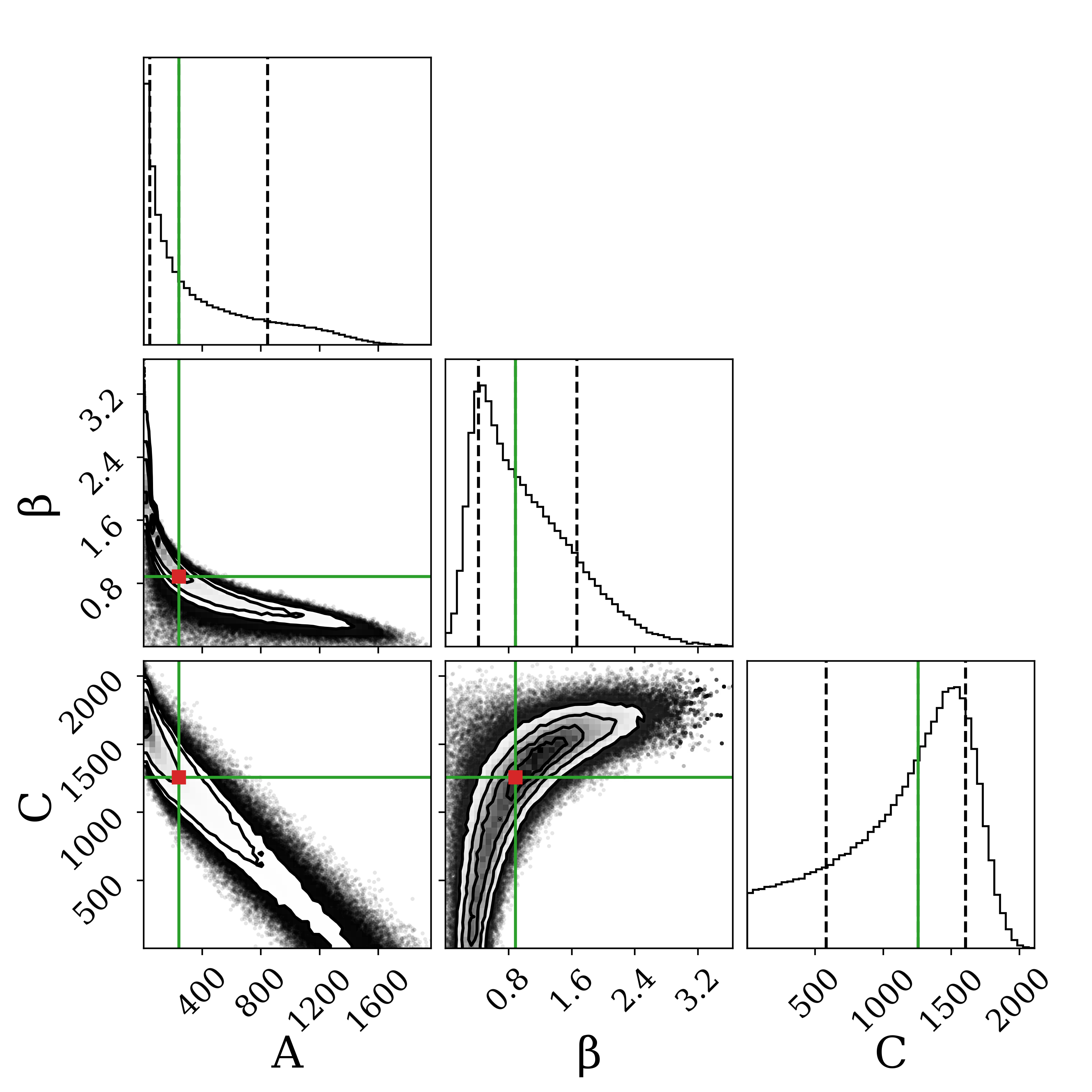}
    \includegraphics[width = 0.5\textwidth]{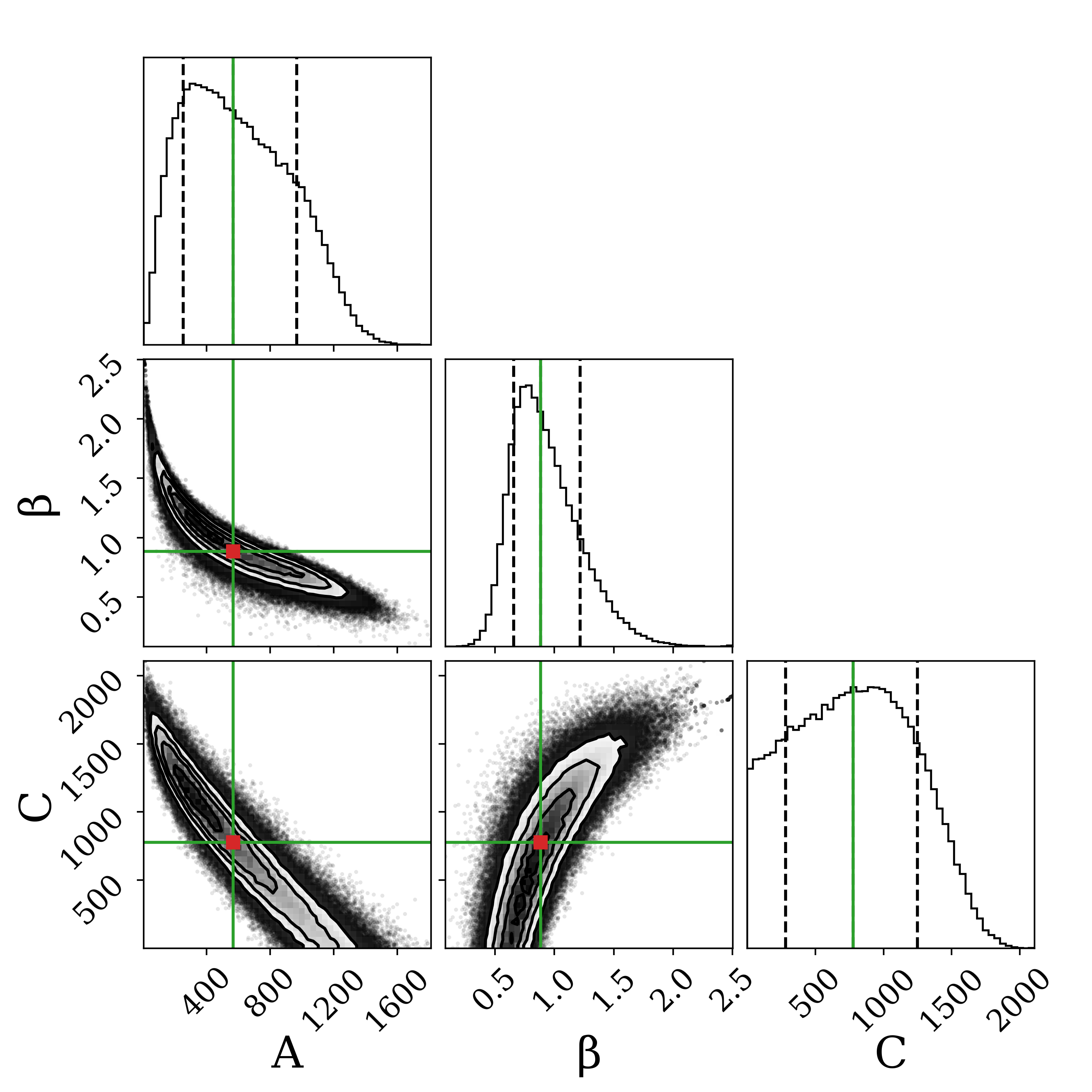}
    \caption{Same as figure~\ref{fig:corner1}, with $\alpha=357.0^{\circ}$ on the left and $\alpha=1.0^{\circ}$ on the right. }
    \label{fig:corner2}
\end{figure*}

\begin{figure*}
    \hspace{-1cm}
    \includegraphics[width =0.5\textwidth]{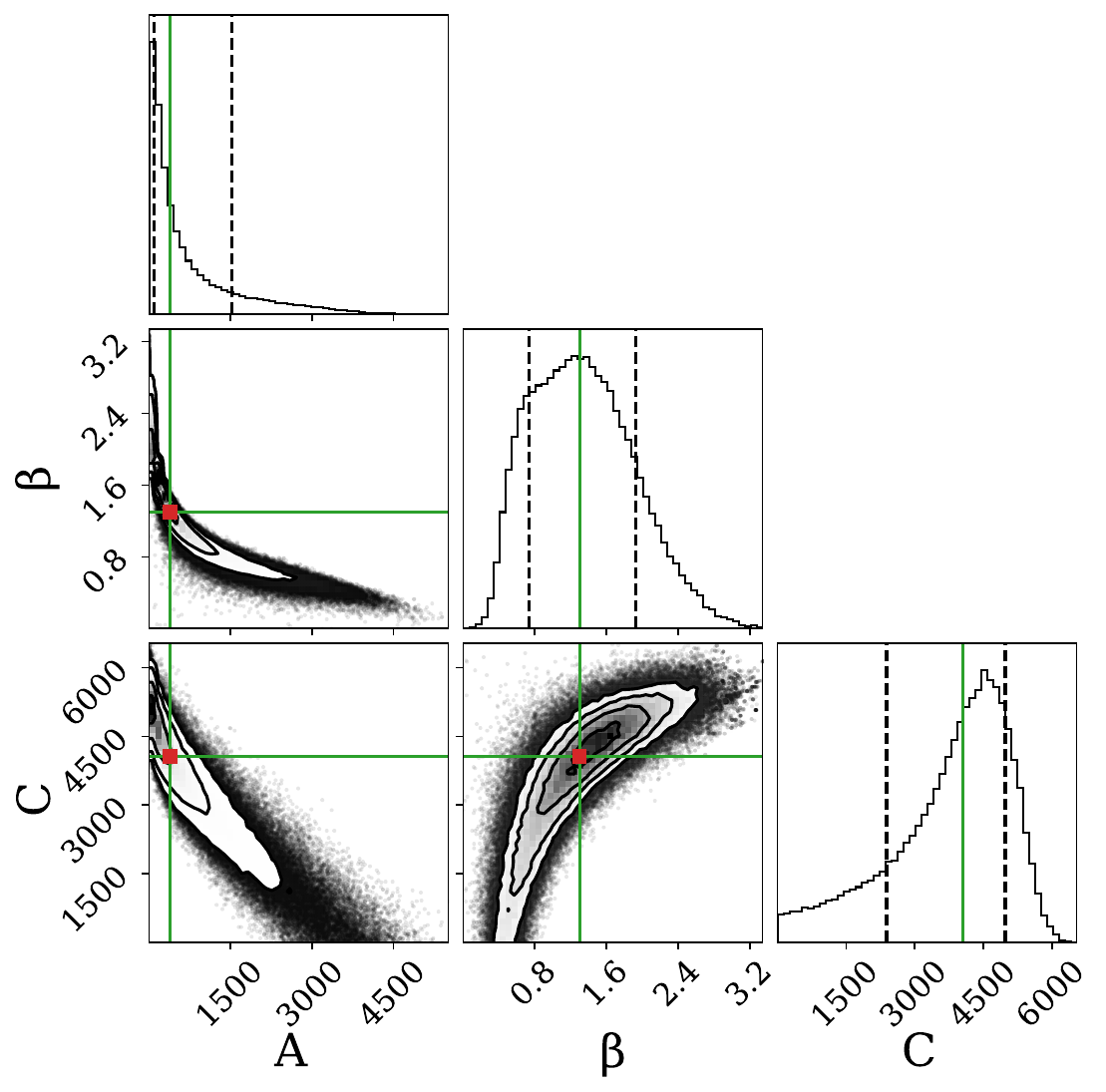}
    \caption{Same as figure~\ref{fig:corner1}, with $\alpha=4.0^{\circ}$.}
    \label{fig:corner3}
\end{figure*}

In figures \ref{fig:corner1} to \ref{fig:corner3}, we show the posterior probability distribution of the model parameters $A,\beta$, and $C$. The diagonal plots show the one-dimensional marginalised posterior distribution of those three parameters, whereas the off-diagonal panels show the posterior of each pair of parameters. The green vertical lines in the diagonal panels in all of the figures show the 50$^{\rm th}$ percentiles of the samples in the marginalised distributions, and two vertical black lines show the 16$^{\rm th}$ and 84$^{\rm th}$ percentiles. The best-fit values (50$^{\rm th}$ percentiles) and their uncertainties (16$^{\rm th}$ and 84$^{\rm th}$ percentiles) of those parameters are shown. We see a strong anti-correlation of parameter $A$ with $\beta$ and $C$. Since we have given only positive prior ranges for all the analysed PCs, we see that in some of the cases the posterior probability distribution is abruptly cut off. Apart from only a few, almost all the posterior distributions are very asymmetrical in nature.

\end{document}